# Exploring Memristive Biosensing Dynamics: A COMSOL Multiphysics Approach

*Manel Bouzouita*
Faculty of Sciences of Monastir
University of Monastir
Monastir, Tunisia
Manel.Bouzouita@ugent.be

*Fakhreddine Zayer*
Center of Autonomous Robotic Systems
Khalifa University
Abu Dhabi, UAE
Fakhreddine.Zayer@ku.ac.ae

*Ioulia Tzovadaki*
Center for Microsystems Technology
Ghent University and IMEC
Technologiepark 126, 9052 Gent,
Belgium
Ioulia.Tzouvadaki@ugent.be

*Sandro Carrara*
Bio/CMOS Interfaces Laboratory
École Polytechnique Fédérale de Lausanne
Neuchâtel, Switzerland
Sandro.carrara@epfl.ch

*Hamdi Belgacem*
Faculty of Sciences Of Monastir
University of Monastir
Monastir, Tunisia
Belgacemhamdi@gmail.com

*Abstract*—This paper presents a novel methodology for modeling memristive biosensing within COMSOL Multiphysics, focusing on critical performance metrics such as antigen-antibody binding concentration and output resistive states. By studying the impact of increasing inlet concentrations, insights into binding concentration curve and output resistance variations are uncovered. The resultant simulation data effectively trains a support vector machine classifier (SVMC), achieving a remarkable accuracy rate of 97%. The incorporation of artificial intelligence (AI) through SVM demonstrates promising strides in advancing AI-based memristive biosensing modeling, potentially elevating their performance standards and applicability across diverse domains.

## I. INTRODUCTION

The integration of memristive devices and biological materials for biosensing applications has emerged as a promising avenue for achieving high sensitivity[1], low power consumption [2], [3], and miniaturization [4], [5]. In the last years, mathematical models became essential for analyzing memristive technologies, paving the way for conducting numerous studies on this topic[6]. In 2017, X. Wang et al.[7] presented a Verilog-A-based memristor model for efficient simulation using voltage-driven state modulation to accurately replicate desired nonlinear attributes.

Nowadays, parallel to the advancement of the memristive biosensing domain, computational modelling has played a crucial role in understanding the relation between biology and electrical characteristics of two-terminal memristive biosensors [8]. In 2015, Tzouvadaki et al. [9] succeeded in building a memristive model illustrating the electrical behaviour obtained experimentally from the nanowire devices before and after the surface bio-modification. Despite the growing interest in memristive biosensing emulation, there is a noticeable gap in the literature concerning numerical modelling of metal oxide memristive (MOM) biosensing systems.

In this study, we present a multiphysics coupled finite element analysis model (FEA) for tantalum oxide-based memristive biosensing. Our contribution lies in elucidating the essential dynamics and parameters governing efficient antigen-antibody binding. Through comprehensive examination of inlet concentration effects, and the resistive behavior, we provide invaluable insights into the mechanisms of memristive biosensing. Additionally, we introduce an SVMC for parallel data processing and categorization. This integration enhances the reading-out process, significantly improving the accuracy and reliability of memristive sensing paradigms. Our proposed model and methodology offer a robust framework for advancing the understanding and application of memristive biosensing technologies.

## II. MEMRISTIVE DEVICE MODELLING

The memristive technology is composed of two tantalum oxide layers and ($Ta^+$) conductive filament (CF) integrated between two palladium electrodes (Pd). Three physical modules are employed to describe the fundamental phenomena inherent within the core structure of the device. At 300K, the MOM model is simulated to demonstrate the resistive switching behaviour according to the study of Sungho Kim et al. [10].

First, The transport of diluted species interface introduces the transmission of oxygen vacancies (OVs) in the device to imitate the construction and rupture of the CF [11]. In this context, A mass balance equation (1) models the diffusion of OVs according to the Fick diffusion effect. The accumulation of OVs forms the CF resulting in the set occurrence. Contrarily, the depletion launches the reset.

$$(\delta c_i/\delta t) + \nabla j + u \nabla c_i = R_i \quad (1)$$

The concentration of oxygen is presented by $c_i$ (mol/m$^3$), the reaction rate is $R_i$ (mol/(m3·s)), u refers to the mass averaged velocity vector (m/s), and $i$ is the mass diffusive flux vector (mol/(m2·s)).

The second interface is an electric current module which computes the electric field and the current distribution based on Ohm's law. A continuity equation (2) is employed to elucidate the computing of the resistive system's output memristance.

$$\nabla \sigma \nabla \psi = 0 \quad (2)$$

$j$ and $\psi$ refer respectively to the current density (A/m2), and the electric potential (V) while $\sigma$ presents the electrical conductivity of the device components. For the CF and the oxide layer Ta2o5, $\sigma$ is illustrated with equation (3). The pre-exponential coefficient $\sigma_0$ is assumed in line with [10] and [12].

$$\sigma = \sigma_0 \exp(-E_{ac}/K*T) \quad (3)$$

The third interface is electromagnetic heating tracing the resistive heating in agreement with joule effect. Indeed, the mobility of the electrons during the CF creation and destruction leads to indirect heat production. This phenomenon is explained with an elliptic partial differential equation (4).

$$(\rho C_p \delta T/\delta t) + \nabla(-k_{th} \nabla T) = Q \quad (4)$$

and $C_p$ interpret respectively the density (kg/m3) and the specific heat capacity at constant pressure (J/(kg·K)), and Q denotes the heat source (W/m3). The thermal conductivity of the Ta2O5 layer and the CF are demonstrated with equation (5). The pre-exponential coefficient $K_{th0}$ is assumed consistently with [10] and [12].

$$K_{th} = K_{th0}(1 + \lambda(T - T_0)) \quad (5)$$

TABLE I
THE GEOMETRICAL AND MATERIAL PARAMETERS OF THE MOM MODEL

| Memristive Structure Components | Parameters | | | |
|---|---|---|---|---|
| | Geometrical | | Material | |
| | *Height (nm)* | *Weight(nm)* | *σ(S/m)* | *K(1/(m.K))* |
| Pd BE | 35 | 50 | 9.34E6 | 70 |
| Ta$_2$O$_5$ | 5 | 42.5 | Eq1 | Eq2 |
| CF (Ta$^+$) | 5 | 7.5 | Eq1 | Eq2 |
| TaO$_x$ | 35 | 50 | 94000 | 0.12 |
| Pd TE | 30 | 50 | 9.34E6 | 70 |

TABLE II
THE PHYSICAL PARAMETERS OF THE MOM MODEL

| Physical Parameters | Value |
|---|---|
| Temperature T | 300 K |
| Hopping distance a | 0.1 nm |
| Linear thermal coefficient λ | 0.1 K$^{-1}$ |
| Escape attempt frequency f | 1E12 Hz |
| Diffusion barrier Ea | 0.85 eV |
| Diffusion coefficient pre-exponential | (1/2)*(a$^2$)* $f$ |

The distinguished parameters of the memristive model are presented in tables I and II.

Fig. 1 shows the variation of logarithmic current as a function of input voltage. The curve is characterized by pinched hysteresis passing through the origins which unravels the typical memory behaviour [13]. Furthermore, The alternation of the voltage induces the resistance switching between the high state (HRS) and the low state (LRS) [14]. In summary, the observed memory effect and the resistive switching behaviour present the ground elements for the evolution of the memristor model into a memristive biosensing framework.

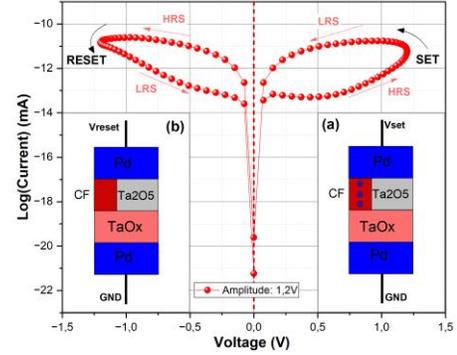

Fig. 1. Variation of the logarithmic current as a function of the input voltage. (**a**): Memristive Low Resistive State (LRS). (**b**): High Resistive State (HRS).

### III. MEMRISTIVE BIOSENSOR MODEL

In this study, we define the analytes as antigens and the receptors as antibodies referring to hajji et al. work [15]. A solution bulk is introduced on the memristive Pd top electrode (TE) in which we integrate the analytes' solution and define the basics of the sensing approach. Under these circumstances, three main physical interfaces are employed. First, a laminar flow interface is used to induct the solution flow according to Navier-Stokes continuity equation (6) for mass and momentum conservation.

$$(\rho \delta u/\delta t) + \rho(u\nabla)u = \nabla[-pI + K] + F \quad (6)$$

p is the pressure (Pa), K is the viscous stress tensor (Pa), and F unveils the volume force vector (N/m3). Second, under transport of the diluted species interface, the mass balance equation (1) is applied in this case for modelling the transport of the antigens from the outside of the bulk to the reaction surface site [16]. Third, we added a surface reaction interface aiming to simulate a simple numerical biosensing model. Thus, a monolayer Langmuir adsorption isotherm [17] is implemented by defining the rate of equation (1) as $R_{i,s}$ and it is solved with equation (7).

$$R_{i,s} = K_{ads}C_{as}(C_{b0} - C_{ab}) - K_{des}C_{ab} \quad (7)$$

The reversible chemical reaction of the adsorption is illustrated as equation (8) where "A" presents the antigen species, "B" indicates the surface-fixed antibodies and AB is the antigen-antibody binding complex. Besides, $K_{ads}$ and $K_{des}$ are, respectively, the Langmuir adsorption and desorption coefficients[15].

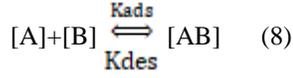

$$[A]+[B] \underset{K_{des}}{\overset{K_{ads}}{\rightleftharpoons}} [AB] \quad (8)$$

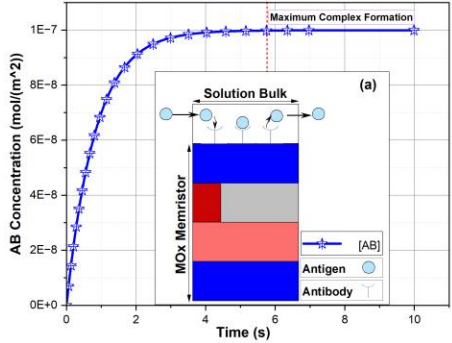

Fig. 2. Variation of complex antigen-antibody concentration as a function of time. **(a)**: Memristive biosensing model structure.

Fig.2 shows the variation of the binding concentration over time. We observe an elevation of the concentration from zero to 1E-7 mol/m^2 for a simulation time equal to 10s. This increase explains the successful attachment of the antigens to fixed antibodies.[15] and [16] proved that the slope of the transition phase is sensitive to various parameters, specifically, the inlet concentration C0 which we report its effect on the resistive state of the memristive device.

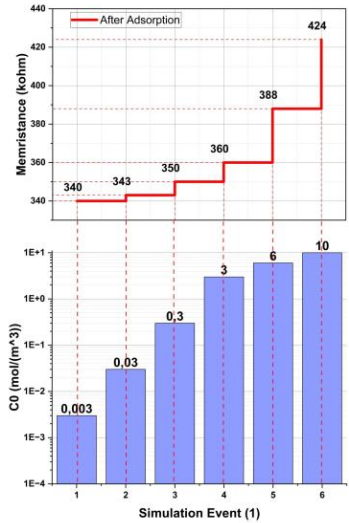

Fig. 3. Variation of resistance state as a function of inlet concentration C0.

Fig.3 displays the variation of the resistive state change as a function of C0. Indeed, before the adsorption, the initial resistive state of the memristive model is equal to 337 Kohm. Then, for a total of six event simulations, we varied the inlet concentration between 0.003 mol/m^3 and 10 mol/m^3. As it is shown in the graph, we notice an increase in the system's generated resistance from 341 Kohm to 421 Kohm. We conclude that this behaviour is explained as typical for a vertical memristor-based biosensor [18]. In brief, our primary significant outcomes pave the way for further enhancement and exploitation. The obtained data is extracted from Comsol Multiphysics and applied for AI implementation as indicated in the next section.

## IV. DATA MAPPING AND CLASSIFICATION

The binding concentration data extracted from Comsol Multiphysics serves as a dataset for proof of concept of machine learning (ML) application. It includes the variation of the concentration and the simulation time as inputs and a classification label as an output. The data undergoes an initial quality-improving cleaning process. Then, the ML algorithm is applied to categorize the data into two labels depending on the concentration phase, specifically, moderate level and high level. In fact, we suppose that the maximum complex formation is the optimum case. Hence, the first half of the transition phase is presumed as a moderate level of antibodies-antigen binding. In opposition, the second remaining range is appraised as a high level. A sample of 38 data is used for primary training. Fig. 4 shows the confusion matrix of an SVMC with a final accuracy rate equal to 97%. Truly, 37 out of the 38 data samples are successfully classified. We observe that 20 of moderate concentrations are predicted correctly with an accuracy of 100%. In brief, this supervised ML classification application is a promising approach towards AI-advanced memristive biosensors, specifically for performance enhancement.

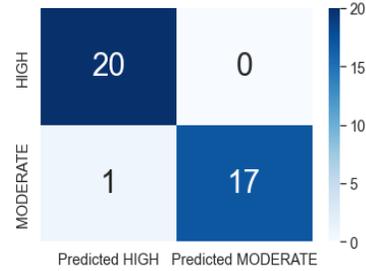

Fig. 4. Confusion matrix of SVM classifying application

## V. CONCLUSION

This study utilized COMSOL Multiphysics to simulate a numerical memristive biosensing model, focusing on the impact of inlet concentration on antigen-antibody binding and memristive resistive states. Through the observation of model behavior, valuable insights into memristive biosensing dynamics were achieved. Additionally, training a proof of concept of SVMC with simulated data achieved an effective 97% accuracy rate, showcasing the potential of AI for advancing the memristive biosensing modeling and performance. In future work, the scope of our simulations is going to be expanded to address the remaining challenges, including enhancing memristor overall performance, improving COMSOL modeling, and ensuring signal compatibility. Overall, continued research in this area promises further development and application of memristive biosensing technologies.